\documentclass[12pt]{article}

\usepackage{epsfig}

\newcommand{\be}{\begin{equation}}
\newcommand{\ee}{\end{equation}}
\newcommand{\bea}{\begin{eqnarray}}
\newcommand{\eea}{\end{eqnarray}}

\begin{document}
\title{
\begin{flushright}
{\small SMI-5-99 }
\end{flushright}
\vspace{1cm} On Stable Sector in Supermembrane Matrix Model}

\author{
I. Ya. Aref'eva${}^{\S}$, A.S. Koshelev${}^{\dag}$ and P. B.
Medvedev${}^{\star}$\\
\\${}^{\S}$ {\it  Steklov Mathematical Institute,}\\ {\it Gubkin st.8,
Moscow, Russia, 117966}\\ arefeva@genesis.mi.ras.ru\\
\\${}^{\star}$
{\it Institute of Theoretical and Experimental Physics,}\\ {\it
B.Cheremushkinskaya st.25, Moscow, 117218}\\
medvedev@heron.itep.ru\\
\\${}^{\dag}$
{\it Physical Department, Moscow State University, }\\ {\it
Moscow, Russia, 119899} \\ kas@depni.npi.msu.su }

\date {~}
\maketitle
\begin{abstract}
We study the spectrum of $SU(2)\times SO(2)$  matrix
supersymmetric quantum mechanics. We use angular coordinates that
allow us to find an explicit solution of the Gauss law constrains
and single out the quantum number $n$ (the Lorentz angular
momentum). Energy levels are four-fold degenerate with respect to
$n$ and are labeled  by $n_q$, the largest $n$ in a quartet. The
Schr\"odinger equation is reduced to two different systems of
two-dimensional partial differential equations. The choice of a
system is governed by $n_q$. We present the asymptotic solutions
for the systems deriving thereby the asymptotic formula for the
spectrum. Odd $n_q$ are forbidden, for even $n_q$ the spectrum has
a continuous part as well as a discrete one, meanwhile for
half-integer $n_q$ the spectrum is purely discrete.
 Taking
half-integer $n_q$ one can cure the model from instability  caused
by the presence of continuous spectrum.
\end {abstract}

\newpage
\section{Introduction}

Supersymmetric quantum mechanics describing an
$SO(d)$ invariant interaction of $d$ $SU(n)$-matrices
has attracted a lot of attention in the last two decades.

The original interest to this model was caused by its description
of $1+0$ reduction of $1+d$ dimensional supersymmetric $SU(n)$
gauge theory. A reduction of non-supersymmetric $d=3$ Yang-Mills
theory to a mechanical model was considered in the pioneering
papers \cite{BMS}. It was noticed that already for the simplest
case of $SU(2)$ gauge group  one gets a rather complicated
mechanical system with 9 degrees of freedom
\cite{Sav}. This mechanical system was investigated only within
the special ansatze. Within one of them the model is reduced to a
model with two degrees of freedom which is still rather
non-trivial and exhibits a chaotic behaviour \cite{BMS,CSS} in the
classical case . Later on, this model \footnote{Nowadays, the
supersymmetric version of this model is known as a toy model and
it is often considered to check and clarify new methods.} has been
investigated in the quantum case. It was also proved that this
system has a discrete spectrum
 \cite{Lus,Sim}.
This spectrum  possesses a very interesting property -- it is in
some sense a direct product of two harmonic oscillators
\cite{BVM}. The possible applications for realistic models  see
\cite{Bad}.

In the end of 80-th the interest to the matrix quantum mechanics
was inspired by an observation that it describes a regularised
membrane theory in $d+2$ space-time dimensions
\cite{Gol,Hop1,WHN,WLN}. The membrane theory was supposed to be
reached in the limit of large $n$. Since the membranes were
considered within the 11 dimensional supergravity the
supersymmetric version of $SU(n)\times SO(9)$ quantum mechanics
has to be examined. In contrast to the bosonic  matrix models
where the spectrum is purely discrete, in supermembrane matrix
models a continuous spectrum, filling the positive half of the
real line was detected \cite{WLN}. This fact was considered as a
manifestation of the instability of the supermembrane against
deformations into stringlike configurations.

Let us also note that more early, in the beginning of the 80's
supersymmetric quantum mechanics (SQM) was proposed  as a  model
for better understanding of supersymmetry breaking  \cite{WSQM}.
In this context SQM was considered in \cite{CH,Flu,BRR}. This
model is essentially simpler as compare to the membrane super
quantum mechanics, since its bosonic part has only one degree of
freedom.

A renovation of interest to a supersymmetric matrix quantum
mechanics in the last three years was motivated by its relation to
a description of the dynamics of D-0 branes in superstring theory
\cite{HT,Wit}.  Moreover, this model in the large $n$ limit
pretends to the role of M-theory \cite{BFSS}. This conjecture has
stipulated  the recent study of  $SU(n)\times SO(d)$
supersymmetric quantum mechanics \cite{DFS}-\cite{Dou}. Within
M-theory there is a very important question of the existence of
normalized eigenfunctions with zero energy, since the zero modes
represent the graviton multiplet of eleven dimensional
supergravity. This problem has attracted attention since the first
paper where the model was introduced. One expected that
$SU(n)\times SO(d)$  supersymmetric quantum mechanics has one
normalized zero-mode for $d=9$, and has none for $d=2,3,5$ (only
in these dimensions a supersymmetric model can be formulated)
\cite{DFS}-\cite{SU3}.

The case of large $n$ is rather involved. The simplest case is the
$SU(2)$ one. In this context the $SU(2)$ quantum mechanics was
investigated in \cite{DFS,KP,Dou}. The  case of arbitrary $n$ was
considered in \cite{HY1,HS,Kon} and $n=3$ was considered in
\cite{SU3}. To investigate the zero-mode  problem the
Born-Oppenheimer approximation was applied \cite{DFS,KP,Dou}. This
method allows to find asymptotic behaviour near "infinity".
Therefore, if one does not expect any singularities at a finite
region, one can deduce an existence/nonexistence of normalized
zero-mode from asymptotic behaviour. An effective  tool to study
the zero-mode problem is an investigation of  a system of first
order differential equations caused by the supersymmetry of a
desired zero mode. The Born-Oppenheimer approximation is also
applicable to the study of the first order differential equations.
Recently the authors of \cite{FGHHY} have performed an
investigation of zero-mode problem using the first order
differential equation. The asymptotic behaviour found in
\cite{FGHHY} supports a common belief about the existence of
normalized zero-mode in $d=9$ and nonexistence in other
dimensions.

Let us make few comments about chaotic behaviour of matrix models.
About classical dynamics of two-dimensional model
\cite{BMS} see
\cite{Med2,Med3}. Classical  dynamics in bosonic membrane matrix
model was investigated and a chaotic behaviour was demonstrated.
Later on, in  $SU(2)\times SO(2)$ matrix model a classical
chaos-order transition was found  \cite{AMRV,AKM}. For the Lorentz
momentum $N$ small enough (even for small coupling constant) the
system exhibits a chaotic behaviour, for $N$ large enough the
system is regular. Up to now the question of similar transition in
quantum case remains open. $\alpha '$-corrections to the
Yang-Mills approximation of D-particle dynamics were studied in
\cite{AFK} where a stabilization of the classical trajectories was
shown.

The main task of this paper is to find the whole spectrum of
$SU(2)\times SO(2)$ supersymmetric matrix quantum mechanics. In
the previous investigations of this model the following particular
results about the spectrum were obtained.  Continuous spectrum was
observed
\cite{WLN}  and the nonexistence of normalized zero mode has been proved
\cite{FH}. The character of spectrum plays an important role in
stability/instability of the system. Let us remind that according
to the commonly accepted opinion this model is unstable. The
potential instability of (super) matrix quantum mechanics is
evident from classical consideration. Namely, the potential of
matrix models has valleys through which a part of coordinates can
escape to infinity without increasing the energy. For the bosonic
case the classical instability is cured by quantum fluctuations
due to which the flat directions become closed by confined
potentials, so that finite energy wave functions fall off rapidly
and spectrum is purely discrete \cite{BVM,Lus}, that provides
stability. The lost of stability  in supersimmetric case is caused
by  additional contributions to the potential coming from the
fermionic degrees of freedom which cancel the bosonic ones. As a
result of this cancellation wave functions are no more confined
and the spectrum becomes continuous \cite{WLN}. This cancellation
takes place on special states, that  means a coexistence of the
continuous spectrum and the discrete one. To specify states for
which a cancellation/non cancellation takes place it is convenient
to arrange  the  states into quartets enumerated by a number
$n_q$. Upon quantization $n_q$ can be integer or half-integer. Our
analysis of $SU(2)\times SO(2)$ supersymmetric quantum mechanics
shows that there is a cancellation in the sector with even $n_q$
and there is no cancellation for half-integer $n_q$ (states with
odd $n_q$ are forbidden). As a result, there is only a discrete
spectrum in the half-integer $n_q$ sector and the model is stable.
The lowest energy in this sector is positive, that means the
supersymmetry is broken in this sector.

Our main tool in the detailed study of the spectral problem of
$SU(2)\times SO(2)$ supersymmetric quantum mechanics is the proper
 coordinates, four angles $\gamma_1$ ,$\gamma_2$, $\alpha$,
$\theta$ and two radii $f$ and $g$. These  coordinates have been
already used in \cite{AMRV}.
 In these coordinates we will
find an explicit solution to the Gauss law constrains and single
out  quantum number $n$.
 Due to this parametrisation the spectral
problem for a supersymmetric version of the model with 6 degrees
of freedom will be reduced to a pair of systems of partial
differential equations in  $f$ and $g$. The choice of a system is
dictated by the value of $n_q$.
 For half-integer $n_q$ we have just one equation on one
function of  $f$ and $g$. For even $n_q$ we have three equations
on three functions of $f$ and $g$.

We present  asymptotic solutions of these two systems deriving
thereby the asymptotic formula for the spectrum. For half-integer
$n_q$ we deal with differential operator corresponding to standard
potential problem in quantum mechanics and  character of the
spectrum can be understood from the form of potential. Since we
have a confining potential the spectrum is discrete. We present an
asymptotic solution of the Schr\"odinger equation and
corresponding formula for the spectrum. For even
$n_q$ we have a matrix second order differential operator. Just
for this system a vanishing  of the bosonic confining potential
takes place on special states and the matrix differential operator
has the continuous spectrum. To get it we find the asymptotic
solution of the system of three equations. Besides the continuous
spectrum this operator possesses the discrete spectrum. States
with integer (half-integer) $n$ belong to the stable sector if
they satisfy the constraint $Q\Psi_n=0$ ($\bar{Q}\Psi_n=0$).

The paper is organised as follows. In Section 2 we explore the
algebraic structure of $SU(2)\times SO(2)$ supersymmetric quantum
mechanics in context of energy level degeneracy. In Section 3 we
specify the angular $SU(2)\times SO(2)$ parametrization. A special
attention is spared to generalized periodicity. In Section 4 we
solve the constrains in the angular parametrization and present
the spectral problem as two sets of two-dimensional partial
differential equations. In Section 5 the spectrum of corresponding
differential operators is examined. We present asymptotic
solutions of the Schr\"odinger equation and corresponding formula
for the spectrum.

\section{Algebraic Structure of $SU(2)\times SO(2)$ Supersymmetric Quantum
Mechanics}

\subsection {The Hamiltonian and Superalgebra}

We consider the system described by the Hamiltonian
\bea
H&=&H_B +H_F ,\label{system}\\ H_B&=&\frac{g_s}2(\pi_1^2+\pi_2^2)+
\frac1{2g_s}\left|\varphi_1\times\varphi_2\right|^2 ,\nonumber\\
H_F &=& -\frac
i2\varepsilon^{abc}\left(\varphi_1^a+i\varphi_2^a\right)\chi^b\chi^c-
\frac
i2\varepsilon^{abc}\left(\varphi_1^a-i\varphi_2^a\right)\bar{\chi}^b\bar{\chi}^c
\nonumber
\eea
and constrained by the Gauss law $j^a|\Psi\rangle=0$.
$\pi_i^a$, $\varphi_j^b$ and
$\bar{\chi}^a$, $\chi^b$ are canonically conjugated pairs with
\be
\left[\pi_i^a,\varphi_j^b\right]=-i\delta^{ab}\delta_{ij},~~~
\left\{\bar{\chi}^a\chi^b\right\}=\delta^{ab} .
\label{comm}
\ee
The  Hamiltonian (\ref{system}) is invariant under $SU(2)$
rotations and redundant Lorentz rotation generated by
\bea
j^a=l^a +s^a ,&& l^a =
\varepsilon^{abc}\varphi_i^b\pi_i^c ,~~
s^a =-i\varepsilon^{abc}\bar{\chi}^b\chi^c ,\\
N= N_B +N_F ,&& N_B =\varphi_1^a\pi_2^a-\varphi_2^a\pi_1^a ,~~
N_F =\frac12\bar{\chi}^a\chi^a
\eea
respectively. The commutation relations for the currents read:
$$
\left[j^a,j^b\right]=i\varepsilon^{abc}j^c ,~~
\left[j^a,N\right]=0.
$$
There are two supercharges:
\bea
Q=-\frac{\sqrt{g_s}}{\sqrt{2}}\bar{\chi}^a(\pi_1^a-i\pi_2^a)+\frac
i{\sqrt{2g_s}}\chi^a\varepsilon^{abc}\varphi_1^b\varphi_2^c ,\\
\bar{Q}=-\frac{\sqrt{g_s}}{\sqrt{2}}\chi^a(\pi_1^a+i\pi_2^a)-\frac
i{\sqrt{2g_s}}\bar{\chi}^a\varepsilon^{abc}\varphi_1^b\varphi_2^c\\
\eea
and
$$
\left\{Q,\bar{Q}\right\}=H .
$$
The supercharges commute with the Hamiltonian up to the generator of $SU(2)$ rotations
vanishing on the physical states.

The remaining commutators with $Q$ and $\bar{Q}$ are the following:
\bea
Q^2&=&0,~~~\bar{Q}^2=0 ,\label{bQalg}\\
\left[Q,j^a\right]&=&\left[\bar{Q},j^a\right]=0 ,\nonumber\\
\left[N,Q\right]&=&-\frac 12 Q ,\nonumber\\
\left[N,\bar{Q}\right]&=&\frac 12 \bar{Q} ,\nonumber\\
\left[N,H\right]&=&0 .
\label{eQalg}
\eea
Note, that the algebra (\ref{bQalg})-(\ref{eQalg}) is not affected
by a shift of $N$ by a constant. As compare with \cite{FGHHY} our
$N$ is shifted on $1/4$.

\subsection{Energy Level Degeneration}

\subsubsection{The Algebra}

On the physical states ($j^a|\Psi \rangle=0$) there are three
operators commuting with $H$: $Q$, $\bar Q$ and $N$. This provides
the degeneration of energy levels. In this section we discuss the
irreducible representations $v_E$ of the algebra (\ref{bQalg}) -
(\ref{eQalg}) for a non-zero energy $E$.

{\bf Lemma.} Let $|\Psi\rangle \in v_E$ with $E\neq 0$ be an eigenvector of $H$ and $N$, then
$Q|\Psi\rangle =0$ or $\bar Q|\Psi\rangle =0$.

{\bf Proof.} Suppose that $Q|\Psi\rangle \neq 0$ and $\bar
Q|\Psi\rangle \neq 0$ then one has two additional eigenvectors of
$N$ with the same energy:
\bea
N(Q|\Psi\rangle )&=&(n-\frac12)(Q|\Psi \rangle ) ,
\label{nq}\\
N(\bar Q |\Psi\rangle )&=&(n+\frac12)(\bar Q |\Psi \rangle ).
\label{n}
\eea One can check, that the operator
$$
 K=HN+\frac1{4}[Q,\bar{Q}]
$$
  is one more Casimir of the algebra (\ref{bQalg}) - (\ref{eQalg}).
By a simple calculation one finds:
 \bea
 K|\Psi\rangle&=&
E(n+\frac14)|\Psi\rangle -\frac12 \bar Q Q |\Psi\rangle=
E(n-\frac14)|\Psi\rangle +\frac12 Q \bar Q |\Psi\rangle ,\label{bCas}\\
 K(Q|\Psi\rangle)& =&
E(n-\frac14)(Q|\Psi\rangle) , \\
 K(\bar Q |\Psi\rangle)&=&
E(n+\frac14)(\bar Q|\Psi\rangle)
\label{eCas}.
\eea
For $v_E$ to be
irreducible the value of the Casimir on $|\Psi\rangle$ should be
equal to its value on $Q|\Psi\rangle$ and $\bar{Q}|\Psi\rangle$.
$~$From equations (\ref{bCas})-(\ref{eCas}) we see that this
requires
\footnote{Remind that if $Q| \Psi\rangle =0$ then $|
\Psi\rangle =Q|\chi\rangle$ with $|\chi\rangle=\frac
1E\bar{Q}|\Psi\rangle$.}
$$
Q|\Psi \rangle =0~~~\mbox{or}~~~\bar{Q}|\Psi\rangle =0.
$$

Since $\bar Q$ and $Q$ act as raising and lowering operators (eqs.(\ref{nq}),
(\ref{n})) one
can introduce a label:  $v_{E,n}$ where $n$ is the value of $N$ on the
vector $|\Psi\rangle \in \ker \bar Q$.

\begin{center}
\epsfig{file=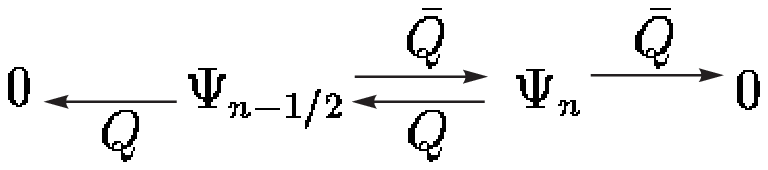,
   width=150pt,
  angle=0
 }
\end{center}

So, we find that $v_{E,n}$
is a vector space spanned by two eigenvectors of $N$:
$|\Psi_1 \rangle$ and $|\Psi_2 \rangle$
such that
$$
\bar{Q}|\Psi_1 \rangle=0=Q|\Psi_2\rangle
$$
and $$
 N |\Psi_1\rangle = n |\Psi_1 \rangle ,~~~ N |\Psi_2 \rangle
= (n- \frac12)|\Psi_2\rangle.
$$

\subsubsection {Discrete Symmetry}

In addition to the algebra (\ref{bQalg}) - (\ref{eQalg}) the
Hamiltonian (\ref{system}) admits one more symmetry. The bosonic
part of the Hamiltonian is invariant under the discrete
transformation
\be
\varphi_1\longleftrightarrow\varphi_2 .
\label{bosdiscr}
\ee
To extend this symmetry for the supersymmetric case we put:
 \bea
\chi\to\mu\bar{\chi}\\
\bar{\chi}\to\nu\chi.
\label{susydiscr}
\eea
The invariance of the Hamiltonian and of the commutation relations
gives the following restrictions on $\mu ,\nu$:
 \bea
-\mu^2=\nu^2=i ,\\ \mu\nu=1.
\eea
One more restriction:
$$
\nu=i\mu
$$
 comes from imposing the condition for transformation
law of supercharges $Q, \bar Q$ to be homogeneous.

The system of equations for $\mu$ and $\nu$
has a solution (up to an overall sign):
$$
\mu=e^{-i\frac{\pi}4},~~~\nu=e^{i\frac{\pi}4}.
$$
Denote the operator generating this symmetry by $P$, then we have
\bea
&[P,H]=0,~~~[P,j]=0, &\label{bPalg}\\
&PQ-\mu\bar{Q}P=0,~~~P\bar{Q}-\nu Q P=0 ,&\nonumber\\
&{}\{P,N\}=\frac32 P ,&\nonumber\\
& P^2=1 .&\label{ePalg}
\eea

Note, that $P$ acts on the spinor monomials without permutations of spinor
factors, because these permutations are incompatible with the invariance of the
anticommutation relations (\ref{comm}).

\begin{figure}[h]
\begin{center}
\epsfig{file=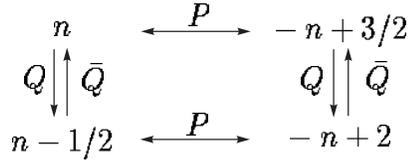,
   width=150pt,
  angle=0,
 }
\caption{States in a quartet $n_q=n$}
\end{center}
\end{figure}

As $Pv_{E,n}=v_{E,2-n}$ (action of $P$-operator on the states will
be specified below) the irreducible representations of the
enlarged algebra
(\ref{bQalg})-(\ref{eQalg})+(\ref{bPalg})-(\ref{ePalg}) are direct
sums: $V_{E,n_q}=v_{E,n}+v_{E,2-n}$, where $n_q$ is the maximal
number of $n$ and $2-n$. To avoid doublecounting one has to
restrict the range for $n$ as to exclude $n<-n+2$ which gives
$n>1$. In the exceptional case $n=1$: $v_{E,1} =v_{E,-1+2}$, {\em
i.e.} for $n=1$ one has a two-dimensional instead of a
four-dimensional irreducible representation. The further
specification of the range will be given below.

\subsection{Action of $P$-operator on the States}

A general wave function is a vector of the eight-dimensional fermionic
Fock space based on the fermionic vacuum $|0\rangle$: $ \chi^a |0\rangle =0$.
It is useful to arrange the fermionic states according to their parity as
follows:
\be
|\Psi\rangle=\left(\psi_0+
\psi^1\bar{\chi}^2 \bar{\chi}^3+
\psi^2\bar{\chi}^3 \bar{\chi}^1+
\psi^3\bar{\chi}^1 \bar{\chi}^2+\
\tilde{\psi}_0\bar{\chi}^1 \bar{\chi}^2\bar{\chi}^3+
\tilde{\psi}^1\bar{\chi}^1+
\tilde{\psi}^2\bar{\chi}^2+
\tilde{\psi}^3\bar{\chi}^3\right)
|0\rangle ,
\label{vector}
\ee
where $\psi_0$, $\psi^a$, $\tilde{\psi}_0$ and $\tilde{\psi}^a$
are some functions of $\vec{\varphi}_1$, $\vec{\varphi}_2$. The
Fock space can also be created by operators $\bar{\chi}^a$ acting
on the "dual" vacuum
$$
|0)=\bar\chi^1\bar\chi^2\bar\chi^3|0\rangle .
$$

To define the action of $P$ on $|\Psi\rangle$ one has to specify the
action of $P$ on the vacuum $|0\rangle$ (or on $|0)$). The unique choice
that gives non-degenerate $P$ is:
\be
P|0\rangle=\lambda|0),
\label{Pvac}
\ee
where $\lambda$ is some numerical factor to be fixed by the
constraint $P^2 =1$.

It is instructive to represent $|\Psi\rangle$ as a column of $\psi$-s with
the ordering dictated by eq.(\ref{vector}). Within this notation one can
write:
\be
P^2 |\Psi\rangle=
P^2
\left(
\begin{array}{l}
\psi_0\\
\psi^1\\
\psi^2\\
\psi^3\\
\tilde{\psi}_0\\
\tilde{\psi}^1\\
\tilde{\psi}^2\\
\tilde{\psi}^3
\end{array}
\right)_{(1,2)}
=
\lambda P \left(
\begin{array}{l}
\tilde{\psi}_0(-\nu^3)\\
\tilde{\psi}^1\nu\\
\tilde{\psi}^2\nu\\
\tilde{\psi}^3\nu\\
\psi_0\\
\psi^1(-\nu^2)\\
\psi^2(-\nu^2)\\
\psi^3(-\nu^2)
\end{array}
\right)_{(2,1)}
=
- \lambda^2 \nu^3\left(
\begin{array}{l}
\psi_0\\
\psi^1\\
\psi^2\\
\psi^3\\
\tilde{\psi}_0\\
\tilde{\psi}^1\\
\tilde{\psi}^2\\
\tilde{\psi}^3
\end{array}
\right)_{(1,2)},
\label{col}
\ee
where labels $(1,2)$, $(2,1)$ indicate the order of bosonic arguments
for $\psi$-s.

The constraint $P^2 =1$ results in $-\lambda^2 \nu^3 =1$ and
gives, according to our choice of $\nu$
\footnote{Starting from the
dual vacuum $|0)$ as $P|0)=\rho |0\rangle$ one finds $\rho
=e^{-i\frac{\pi}8}$. It is a matter of simple algebra to check
that for these $\lambda$ and $\rho$ $P$ is correctly defined {\em
i.e.} it does not depend on the choice of the cyclic vector,
$|0\rangle$ or $|0)$, in the fermionic Fock space.}:
$$
\lambda=e^{i\frac{\pi}8}.
$$

In a matrix form the action of the operator $P$ on the fermionic degrees
 of freedom can be represented as follows
\be
P=
 \lambda
 \left(
\begin{array}{cc} 0&
\begin{array}{cccc} 1&0&0&0\\ 0&-\nu^2
&0&0\\ 0&0&-\nu^2 &0\\ 0&0&0&-\nu^2 \\
\end{array}\\
\begin{array}{cccc}
-\nu^3&0&0&0\\
0&\nu &0&0\\
0&0&\nu &0\\
0&0&0&\nu
\end{array}& 0\\
\end{array}
\right).
\ee

\subsection{Gauss Law and $N|\Psi\rangle =n|\Psi\rangle$ in Components}

The Gauss law $$ j^a |\Psi\rangle =0 $$ fixes the $SU(2)$
transformation properties of the component wave functions
$\psi^a$, $\tilde {\psi}^a$. These are given by: $$ l^a
|\Psi\rangle =-s^a |\Psi\rangle . $$ The spin operator $\vec s$
acts on fermionic Fock states and its action can be easily
calculated to give
\begin{eqnarray}
  [l^a ,\psi_0 ]&=&0 , \nonumber  \\
{}[l^a ,\psi^{b} ]&=& i\epsilon^{abc}\psi ^c , \nonumber \\
\label{GL}
{}[l^a ,\tilde \psi_0 ]&=&0 , \\
{}[l^a ,\tilde \psi^b ]&=   &
i\epsilon^{abc}\tilde \psi^c .\nonumber
\end{eqnarray}
Hence, we conclude that: 1) $\psi_0$ and $\tilde \psi_0$ are $SU(2)$
singlets, 2) $\vec \psi =(\psi^1 ,\psi^2, \psi^3)$ and
$\vec{\tilde \psi} =(\tilde \psi^1 ,\tilde \psi^2 ,\tilde  \psi^3)$ are
$SU(2)$ triplets. One can check that
$$
{\vec l}^2 \psi_0 = {\vec l}^2 \tilde \psi_0  =0,
$$
$$
[\vec{ l}^2 , \vec\psi] =2\vec \psi ,~~~~
 [\vec{ l}^2 , \vec{\tilde\psi}] =2\vec{\tilde \psi}
$$ as it must be for the scalar and vector representation. The
explicit solution of (\ref{GL}) will be given below after the
parametrisation of the configuration space $(\vec{\varphi_1}
,\vec{\varphi_2} )$ will be specified.

In the following we shall exploit the eigenstates of $N$. It is
instructive to separate the fermion number operator $N_F$. Let:
$N|\Psi\rangle =n|\Psi\rangle $ then $N_B |\Psi\rangle =(n - N_F )
|\Psi\rangle$  gives
\begin{eqnarray}
N_B \psi_0 &=&n\psi_0 ,\nonumber\\
N_B \vec\psi &=&(n-1)\vec\psi ,
\nonumber \\
N_B \tilde\psi_0 &=&(n-\frac32)\tilde\psi_0 ,
\label{Nb} \\
N_B \vec{\tilde\psi} &=&(n-\frac12)\vec{\tilde\psi}.
\nonumber
\end{eqnarray}

\subsection{Shr\"odinger equation in components}

The fermionic Fock space decomposition (\ref{vector}) for
$|\Psi\rangle$ provides
 a matrix
representation for Shr\"odinger equation $H|\Psi\rangle =E|\Psi\rangle$.
Taking $|\Psi\rangle$ as a column (see (\ref{col})) one deduces $H$ in a
block-diagonal form:
\be
H=
\left(
\begin{array}{cc}
\begin{array}{cccc}
H_B& i\varphi^1&i\varphi^2&i\varphi^3\\
-i\bar \varphi^1&H_B&0&0\\
-i\bar \varphi^2&0&H_B&0\\
-i\bar \varphi^3&0&0&H_B\\
\end{array}& 0 \\
0&
\begin{array}{cccc}
 H_B & - i\bar \varphi^1 & -i\bar \varphi^2 & -i\bar \varphi^3 \\
 i \varphi^1 & H_B & 0 & 0 \\
 i \varphi^2 & 0 & H_B & 0 \\
 i \varphi^3 & 0 & 0 & H_B \\
\end{array}
\end{array}
\right),
\ee
where
\be
\varphi^a =\varphi^a_1 +i\varphi^a_2,~~\bar{\varphi}^a =\varphi^a_1
-i\varphi^a_2.
\label{com}
\ee

On physical states ($\vec j |\Psi\rangle =0$) the component equations for
$SU(2)$ triplet $\vec\psi$ ($\vec{\tilde\psi}$) are not independent.
For instance, the last equation for the upper block is (equation number four):
\be
-i\bar{\varphi}^3 \psi_0 +H_B \psi^3 =E\psi^3 .
\label{schrod4}
\ee
Let us apply $l^1$ to (\ref{schrod4}):
$$
-il^1(\bar{\varphi}^3 \psi_0 ) +l^1(H_B \psi^3 ) =El^1 \psi^3 ,
$$
then by using the commutation relations (\ref{GL}) and $[l^a ,\varphi^b]
=i\varepsilon^{abc} \varphi^c$ one gets the third Shr\"odinger equation
$$
-i\bar{\varphi}^2 \psi_0 +H_B \psi^2 =E\psi^2
$$
and so on.

Therefore, we are left with two sets of equations:
\be
\left\{
\begin{array}{l}
H_B \psi_0 +i(\vec{\varphi} \vec{\psi})
=E\psi_0\\
H_B \psi^3-i\bar{\varphi}^3 \psi_0 =E\psi^3
\end{array} \right.
\label{schrod}
\ee
and
\be
\left\{
\begin{array}{l}
H_B \tilde{\psi}_0 -i(\vec{\bar{\varphi}} \vec{\tilde{\psi}})
=E\tilde{ \psi}_0\\
H_B \tilde{ \psi}^3+i\varphi^3\tilde{ \psi}_0 =E\tilde{ \psi}^3
\end{array} \right.
\label{tschrod}
\ee
with $\psi^{1,2}$ ($\tilde{\psi}^{1,2}$) expressed in terms of $\psi^3$
($\tilde{\psi}^3$).

\section{$SU(2)\times SO(2)$ Parametrisation}

\subsection{Parametrisation}

We parametrise the configuration space $\{\varphi_i^a\}$ by using
the new coordinates $\{f, g, \theta, \gamma_1, \gamma_2, \alpha\}$
as follows :
\bea
\varphi_1=\frac{1}{\sqrt{2}}U^+\left(\sigma^1f\cos\theta-
\sigma^2g\sin\theta\right)U\nonumber\\
\varphi_2=\frac{1}{\sqrt{2}}U^+\left(\sigma^1f\sin\theta+
\sigma^2g\cos\theta\right)U\label{rcs} .
\eea
Here $\sigma^a$
\bea
\sigma^1=\left(
\begin{array}{cc}
0&1\\
1&0
\end{array}
\right),~ \sigma^2=\left(
\begin{array}{cc}
0&-i\\
i&0
\end{array}
\right),~ \sigma^3=\left(
\begin{array}{cc}
1&0\\
0&-1
\end{array}
\right),\nonumber\\
\eea
are Pauli matrices and the $SU(2)$ matrix $U(\gamma_1 ,\gamma_2 ,\alpha)$ is
given by
\bea
U(\gamma_1,\alpha,\gamma_2)=\exp\left(\frac i2\gamma_1\sigma_3\right)
  \exp\left(\frac i2\alpha\sigma_1\right)
  \exp\left(\frac i2\gamma_2\sigma_3\right).
\nonumber
\eea
The range for the new coordinates is
$0\leq\gamma_1,\gamma_2 < 2\pi$, $-\pi<\alpha<0$, $0\leq\theta<\pi/2$, $f, g>0$

\begin{figure}
\begin{center}
\epsfig{file=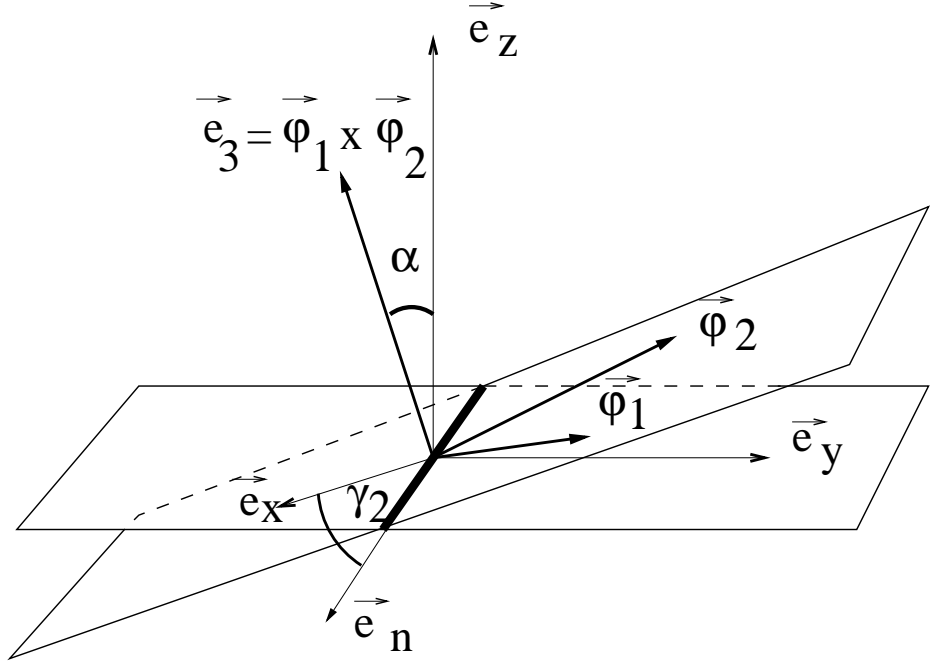,
   width=350pt,
  angle=0
 }
\end{center}
\label{Fig2} \caption{Angular parametrization }
\end{figure}

The $SU(2)$ angular variables $\gamma_1, \alpha, \gamma_2$ are
similar to the Euler angles. The $SU(2)$ adjoins $\varphi_i$ can
be viewed as the $R^3$ vectors $\vec{\varphi}_i$. Denote
$\vec{e}_x , \vec{e}_y , \vec{e}_z$ the unit vectors of the fixed
coordinate system.
$\vec{e}_n=\vec{e}_z\times(\vec{\varphi}_1\times\vec{\varphi}_2)$
is a vector along the knot-line (the line of intersection of the
$(\vec{\varphi}_1, \vec{\varphi}_2)$ and $(x, y)$ planes). The
first rotation on $\gamma_2$, around the $z$-axis matches
$\vec{e}_n$ with $\vec{x}$. After this rotation the vector
$\vec{e}_3=\vec{\varphi}_1\times\vec{\varphi}_2$ falls into the
$(y, z)$ plane, with $(\vec{e}_3)_y<0$. The second rotation around
the $x$-axis in clockwise direction on the angle $\alpha$, matches
$\vec{e}_3$ with $\vec{z}$ that gives $-\pi <\alpha <0$. These two
rotations place the pair $\vec{\varphi}_1, \vec{\varphi}_2$ into
the  $(x, y)$ plane and $\angle (\vec{\varphi}_1, \vec{\varphi}_2)
<\pi$ (
 $(\vec{e}_3)_z$ is positive).

To fix the third angle $\gamma_1$, let us take a pair
$(\vec{\varphi}_1, \vec{\varphi}_2)$ in $(x, y)$ plane with
$|\vec{\varphi}_1| =r_1$ , $|\vec{\varphi}_2| =r_2$  and $\angle
(\vec{\varphi}_1, \vec{\varphi}_2) =\beta$ and examine its orbit
under the action of the rotations around the $z$-axis. This orbit
is the set $\{ (r_1 \cos\delta ,r_1 \sin\delta),(r_2 \cos (\delta
+\beta ) ,r_2 \sin (\delta +\beta )\}$ with $0\le \delta < 2\pi$.
We call a pair $(\vec{\varphi}_1, \vec{\varphi}_2)$ as
"$SO(2)$-orthogonal" if
\be
\varphi_1^1\varphi_1^2+\varphi_2^1\varphi_2^2=0,
 \label{u1cond}
\ee
here $\varphi_i^1=(\vec{\varphi}_i)_x ,
\varphi_i^2=(\vec{\varphi}_i)_y$. The $SO(2)$-orthogonality
condition for the points of the orbit reads:
$$
\tan 2\delta =-\frac{r_2^2 \sin 2\beta}{r_1^2 +r_2^2 \cos 2
\beta} .
$$
Hence, at any orbit there are four $SO(2)$-orthogonal
pairs $(\vec{\varphi}_1, \vec{\varphi}_2)$. One of these pairs has
$\vec{\varphi}_1$ lying in the fourth quadrant {\em i.e.} $\varphi
_1^1
 >0$, $\varphi_1^2 \le 0$. This pair we call the special one. For any pair
$(\vec{\varphi}_1, \vec{\varphi}_2)$ we define $\gamma _1$ as the angle of
rotations around the $\vec{z}$-axis that matches
$(\vec{\varphi}_1, \vec{\varphi}_2)$, with the special pair. The range for
$\gamma_1$ is obviously $0\le \gamma_1 <2\pi $.

\begin{figure}
\begin{center}
\epsfig{file=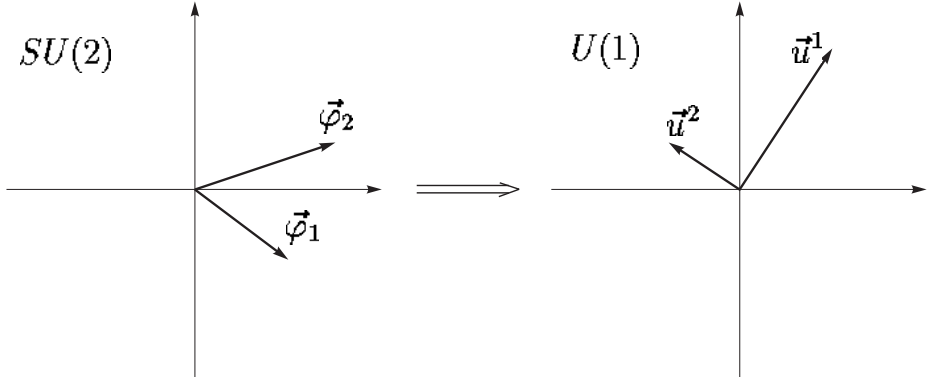,
   width=350pt,
  angle=0
 }
\end{center}
\label{Fig3} \caption{$SU(2)$ and $SO(2)$}
\end{figure}

To parametrise the special pair $(\vec{\varphi}_1,
\vec{\varphi}_2)$ note that the numbers $\varphi^a_i$ form also a
pair of $SO(2)$ (Lorentz) orthogonal
 vectors:  $u^1=\{\varphi_1^1,\varphi_2^1\}$ and
 $u^2=\{\varphi_1^2,\varphi_2^2\}$,
so that $\arg u^2 =\arg u^1 +\frac{\pi}{2}$.
The remaining coordinates $\{f, g, \theta\}$ are defined to be the polar
coordinates for
 $u^1$ and $u^2$:
\be
\begin{array}{l}
\varphi_1^1=f\cos\theta\\
\varphi_2^1=f\sin\theta
\end{array}
~~~
\begin{array}{l}
\varphi_1^2=-g\sin\theta\\
\varphi_2^2=g\cos\theta .
\end{array}
\ee
As the pair is the special one the ranges for $\{f, g, \theta\}$ are
$0\le\theta<\pi/2$  and $f,g >0$.

To summarize, we have proved that by the $SU(2)$ rotation defined
by the Euler-like angles $\gamma_1 $, $\gamma_2$ and $ \alpha$ any
pair $(\vec{\varphi}_1, \vec{\varphi}_2)$ is matched with the
special pair which is parametrised by $f,g, \theta$. The special
case of an angular momentum parametrisation was proposed in
\cite{Bad}.

\subsection{Generalized Periodicity}

To extend the ranges for the angular variables $\gamma_i$ up to
$R^1$ axis we employ the usual $2\pi$ periodicity conditions for
$\gamma_i$. As to the $\theta$ angle with $0\le\theta<\pi/2$ the situation
is more subtle.

Let us take a point with $SO(2)$ angle coordinate outside the
range: $(f,g,\theta +\frac{\pi}{2} ,\gamma_1)$ ($\gamma_2$ and
$\alpha$ are irrelevant for the subsequent discussion). By means
of (\ref{rcs}) this set parametrise the pair
 $$
 \varphi_{1,2} =U^+
(\gamma_1) \chi_{1,2}U(\gamma_1),~~~U(\gamma_1)=e^{\frac
i2\gamma_1\sigma_3}
$$
where
\bea
\chi_1& =&-\sigma_1 f\sin\theta
+ \sigma_2 g\cos\theta  ,\nonumber\\
\chi_2& =&\sigma_1 f\cos\theta
-\sigma_2 g\sin\theta .
\eea
The pair $(\vec{\chi}_1 ,\vec{\chi}_2)$
is $SO(2)$-orthogonal but not special. What are the true
coordinates for $\varphi_{1,2}$?

In complex notation (\ref{com}) these read:
\bea
\varphi&=& \left(
\begin{array}{cc}
e^{\frac{i}2 \gamma_1}&0\\
0&e^{-\frac{i}2 \gamma_1}
\end{array}
\right)^+
e^{i(\theta +\frac\chi 2)}
(\sigma_1 f +i\sigma_2 g)
\left(
\begin{array}{cc}
e^{\frac{i}2 \gamma_1}&0\\
0&e^{-\frac{i}2 \gamma_1}
\end{array}
\right)  \nonumber \\
&=& e^{i \theta}
\left(
\begin{array}{cc}
0&e^{-i(\gamma_1 -\frac{\chi}2)} (f+g) \\
e^{i(\gamma_1 +\frac{\chi}2 )}(f-g) &0
\end{array}
\right)  \nonumber \\
&=& e^{i \theta}
\left(
\begin{array}{cc}
0&e^{-i(\gamma_1 -\frac{\chi}2)} (f+g) \\
e^{i(\gamma_1 -\frac{\chi}2 )}(g-f) &0
\end{array}
\right)  \nonumber
\eea

Hence, the following points
must be identified
\be
(f,g,\theta +\frac{\pi}{2} ,\gamma_1 ,\gamma_2 ,\alpha) \cong
(g,f,\theta ,\gamma_1 -\frac{\pi}{2} ,\gamma_2 ,\alpha) .
\label{pe}
\ee
 Below we shall call (\ref{pe}) the generalized periodicity condition in $\theta$.

\subsection{ $P$ symmetry in $SU(2)\times SO(2)$ Parametrisation}

On bosonic degrees of freedom $P$ operator acts as
$\vec{\varphi}_1 \leftrightarrow \vec{\varphi}_2$. Here we
describe this action in terms of the new coordinates.  Let us
denote $\vec{\varphi}_1 ' =\vec{\varphi}_2$, $\vec{\varphi}_2 '
=\vec{\varphi}_1$, then $P\vec{\varphi}_i =\vec{\varphi}_i  '$. We
also shall indicate by primes all the objects referred to the pair
$(\vec{\varphi}_1 ' ,\vec{\varphi}_2 ')$.

By definition one has $\vec{e}~' _3  =-\vec{e}_3$ and $\vec{e}~'_n
=-\vec{e}_n$ hence, $\gamma_2^{\prime}=-\pi+\gamma_2$. After
$\gamma_2$ rotation $\vec{e}~'_3 $ lies in the $(x,y)$ plane with
$(\vec{e}~'_3 )_z=-(\vec{e}_3)_z$ and $(\vec{e}~'_3 )_y=
(\vec{e}_3)_y$. This yields: $\alpha^{\prime}=-\pi-\alpha$.

Let us compare the results of two
first rotations for both the pairs
$(\vec{\varphi}_1   ,\vec{\varphi}_2 )$ and
$(\vec{\varphi}_1 '  ,\vec{\varphi}_2 ')$
that lie in $(x,y)$ plane
(we still denote the resulting pairs as
$(\vec{\varphi}_1   ,\vec{\varphi}_2 )$,
$(\vec{\varphi}_1 '  ,\vec{\varphi}_2 ')$
). By means of a simple algebra one finds that for $\gamma_2 '$ and
$\alpha '$ given above:
$$
\left(
\begin{array}{ccc}
-1&0&0\\
0&1&0\\
0&0&-1
\end{array}
\right)\times
\left(
\begin{array}{ccc}
1&0&0\\
0&\cos\alpha&\sin\alpha\\
0&-\sin\alpha&\cos\alpha
\end{array}
\right)\times
\left(
\begin{array}{ccc}
\cos\gamma_2&\sin\gamma_2&0\\
-\sin\gamma_2&\cos\gamma_2&0\\
0&0&1
\end{array}
\right)=
$$
\be
\left(
\begin{array}{ccc}
1&0&0\\
0&\cos\alpha^{\prime}&\sin\alpha^{\prime}\\
0&-\sin\alpha^{\prime}&\cos\alpha^{\prime}
\end{array}
\right)\times
\left(
\begin{array}{ccc}
\cos\gamma_2^{\prime}&\sin\gamma_2^{\prime}&0\\
-\sin\gamma_2^{\prime}&\cos\gamma_2^{\prime}&0\\
0&0&1
\end{array}
\right) .
\ee
{\em i.e.} $(\vec{\varphi}_1 '  ,\vec{\varphi}_2 ')$
and $(\vec{\varphi}_1   ,\vec{\varphi}_2 )$, differs by the
rotation on $\pi$ around the $y$-axis. Thus on the $(x,y)$ plane
one has:
\bea
\varphi_1^{\prime 1}=-\varphi_2^1 &~~&
\varphi_1^{\prime 2}= \varphi_2^2  \nonumber\\
\varphi_2^{\prime
1}=-\varphi_1^1 &~~& \varphi_2^{\prime 2}= \varphi_1^2 . \nonumber
\eea
 One can rewrite these relations in polar coordinates (we use
complex notation):
\bea
\mbox{if}~~~~~~~~~~ \vec{\varphi}_1 =r_1
e^{i\mu}, && \vec{\varphi}_2 =r_2 e^{i\nu} \\
 \mbox{then}~~~
\vec{\varphi}~'_1  =r_2 e^{i(\pi -\nu )}, && \vec{\varphi}~'_2
=r_1 e^{i(\pi -\mu )} .
\eea

Let $\gamma_1$ be the angle that matches $(\vec{\varphi}_1 ,\vec{\varphi}_2)$
with the special pair $(r_1 e^{i(\mu +\gamma_1)} ,r_2 e^{i(\nu +\gamma_1)})$.
The pair $(r_2 e^{-i(\nu +\gamma_1)} ,r_1 e^{-i(\mu +\gamma_1)})$ is also
a special one and
$$
(r_2 e^{-i(\nu +\gamma_1)} ,r_1 e^{-i(\mu +\gamma_1)})
=e^{-i(\gamma_1 +\pi)} (r_2 e^{i(\pi -\nu )} ,r_1 e^{i(\pi -\mu
)})
$$
that yields $\gamma_1 ' =-\gamma_1 -\pi =-\gamma_1 +\pi$
due to periodicity in $\gamma_1$. The $f',g'$ and $\theta '$ are
easily found to be: $f '=f$, $g'=g$ and $\theta '= \pi /2
-\theta$.

Collecting all the pieces together we find the action of $P$ in new coordinates:
\bea
\gamma_2&\longrightarrow&-\pi+\gamma_2\label{bnewdiscr} \nonumber\\
\alpha&\longrightarrow&-\pi-\alpha \nonumber\\
\gamma_1&\longrightarrow&\pi-\gamma_1 \nonumber\\
\theta&\longrightarrow&\pi/2-\theta\\
f&\longrightarrow& f \nonumber\\
g&\longrightarrow& g \nonumber
\label{enewdiscr}
\eea

\section{Shr\"odinger equation in $SU(2)\times SO(2)$ parametrisation}

\subsection{$N=n$ and Gauss law -- an explicit solution}

The $SU(2)\times SO(2)$ variables provide the explicit solution
for wave function $|\Psi\rangle$ dependence on angular variables.

For $N_B$ one deduces
$$
N_B=-i\frac{\partial}{\partial\theta} ,
$$
{\em i.e.} $N_B$ eigenfunctions are the plane waves
$e^{ik\theta}$. Applying periodicity condition (\ref{pe}) four
times we find that wave functions components have to be $2\pi$
periodic so $k$ must be an integer. Thus, from eqs.(\ref{Nb}) it
follows that {\em $n$ can be integer or half-integer} and
\be
\mbox{for integer}~n:~~~\Psi_n\equiv
\left(
\begin{array}{rcc}
&\psi_0&\nonumber\\
&\vec{\psi}&\nonumber\\
&0&\nonumber\\
&\vec 0&
\end{array}
\right) ~~~~~~~~
\mbox{for half-integer}~n:~~~\Psi_n\equiv
\left(
\begin{array}{rcc}
&0&\nonumber\\
&\vec 0&\nonumber\\
&\tilde{\psi}_0&\nonumber\\
&\vec{\tilde{\psi}}&
\end{array}
\right) .
 \label{nn}
 \ee
 In the following we shall concentrate on
the case of integer $n$, the half-integer case will be sometimes
discussed for the sake of completeness.

The components of $SU(2)$ generators $l^a$ depend only on "Euler" angles:
\bea
l^1&=&-i\left(\cos\gamma_2\frac{\partial}{\partial\alpha}+
\frac{\sin\gamma_2}{\sin\alpha}\left(\frac{\partial}{\partial\gamma_1}-
\cos\alpha\frac{\partial}{\partial\gamma_2}\right)\right) ,\\
l^2&=&-i\left(\sin\gamma_2\frac{\partial}{\partial\alpha}-
\frac{\cos\gamma_2}{\sin\alpha}\left(\frac{\partial}{\partial\gamma_1}-
\cos\alpha\frac{\partial}{\partial\gamma_2}\right)\right) ,\\
l^3&=&-i\frac{\partial}{\partial\gamma_2} ,
\label{l3new}
\\
(\vec{l})^2&=&-\frac{\partial^2}{\partial\alpha^2}-
\frac{\cos\alpha}{\sin\alpha}\frac{\partial}{\partial\alpha}-
\frac 1{\sin^2\alpha}\left( \frac{\partial^2}{\partial\gamma^2_1}+
\frac{\partial^2}{\partial\gamma^2_2}-
2\cos\alpha\frac{\partial}{\partial\gamma_1}
\frac{\partial}{\partial\gamma_2}\right) .
\label{llnew}
\eea
These expressions coincide with the usual Euler angles
parametrisation for $SO(3)$ generators
\cite{gel}
up to some modification of variables.

For $\psi_0$ ($\tilde{\psi}_0$) eq.(\ref{GL}) shows that $\psi_0$ does not
depend on $(\gamma_1 ,\gamma_2 ,\alpha)$ and has the form:
\bea
\mbox{for integer}~n:~~~&\psi_0 =&e^{in\theta}F_0(f,g)
\nonumber\\
\mbox{for half-integer}~n:~~~&\tilde{\psi}_0=
&e^{i(n-\frac32 )\theta}\tilde{F}_0(f,g).
\eea

From (\ref{l3new}) one finds $\psi^3$ ($\tilde{\psi}^3 $) to
depend only on $\alpha$ and $\gamma_1$. Thus, taking $\psi^3
(f,g,\theta ,\alpha ,\gamma_1 )= \sum F (f,g,\theta )\Phi (\alpha
,\gamma_1 )$ from (\ref{llnew}) one has:
\be
\left(\frac{\partial^2}{\partial\alpha^2}+
\frac{\cos\alpha}{\sin\alpha}\frac{\partial}{\partial\alpha}+
\frac 1{\sin^2\alpha}
\frac{\partial^2}{\partial\gamma^2_1}\right)\Phi =-2\Phi .
\label{l22}
\ee
The solutions of (\ref{l22}) are spherical functions $Y_{1,m}$ with
$m=0,+1,-1$. Taking for $Y_{1,m}$ the explicit expressions we get:
\bea
\mbox{for integer}~n:~~~~~~\psi^3&=&e^{i(n-1)\theta}
\left(F^0\cos\alpha+F^+\sin\alpha e^{i\gamma_1}+F^-\sin\alpha
e^{-i\gamma_1}\right) ,\label{sol3}\\
\\
\mbox{for half-integer}~n:~~\tilde{\psi}^3&=&e^{i(n-1/2)\theta}
\left(\tilde{F}^0\cos\alpha+\tilde{F}^+\sin\alpha e^{i\gamma_1}+
\tilde{F}^-\sin\alpha e^{-i\gamma_1}\right)\label{sol3t}
\eea
where the harmonics $F^0, F^+, F^-$ are functions of $f$ and $g$ only.

We list the remaining components of  $\vec\psi$
($\vec{\tilde{\psi}}$) which are restored by acting of $l^a$ on
$\psi^3$ ($\tilde{\psi}^3  $). It is useful to slightly modify the
basis by passing to the eigenvectors of $l^3$ ($s^3$):
\bea
|\Psi\rangle&=& (\psi_0+
\psi^{-}(\bar{\chi}^1+i\bar{\chi}^2)\bar{\chi}^3+
\psi^{+}(\bar{\chi}^1-i\bar{\chi}^2)\bar{\chi}^3+
\psi^3\bar{\chi}^1\bar{\chi}^2\nonumber\\
& + &\tilde{\psi}_0+
\tilde{\psi}^-(\bar{\chi}^1+i\bar{\chi}^2)+
\tilde{\psi}^+(\bar{\chi}^1-i\bar{\chi}^2)+
\tilde{\psi}^3\bar{\chi}^3)|0\rangle .
 \eea
 In this basis
$\psi^{\pm}$ are achieved by using the raising and lowering
operators $l^{\pm}=l^1\pm il^2$ as
\be
\psi^-=-\frac i2l^-\psi^3,~~~~~
\psi^+=-\frac i2l^+\psi^3 ,
\ee
similarly for the tildes.
 Finally, we have\\
for integer $n$:
\bea
\psi^+ &=&\frac
12e^{i(n-1)\theta}e^{i\gamma_2}
\left(F^0\sin\alpha-F^+(\cos\alpha+1)e^{i\gamma_1}-F^-
(\cos\alpha-1)e^{-i\gamma_1}\right), \label{sol+}\\
 \psi^-
&=&\frac 12e^{i(n-1)\theta}e^{-i\gamma_2}
\left(F^0\sin\alpha-F^+(\cos\alpha-1)e^{i\gamma_1}-F^-
(\cos\alpha+1)e^{-i\gamma_1}\right),
\label{sol-}
 \eea
for half-integer $n$:
\bea
 \tilde{\psi}^+&=&-\frac i2e^{i(n-1/2)\theta}e^{i\gamma_2}
\left(\tilde{F}^0\sin\alpha-\tilde{F}^+(\cos\alpha+1)e^{i\gamma_1}-
\tilde{F}^-(\cos\alpha-1)e^{-i\gamma_1}\right),\\
\tilde{\psi}^-&=&\frac i2e^{i(n-1/2)\theta}e^{-i\gamma_2}
\left(\tilde{F}^0\sin\alpha-\tilde{F}^+(\cos\alpha-1)e^{i\gamma_1}-
\tilde{F}^-(\cos\alpha+1)e^{-i\gamma_1}\right).
\eea

Further restrictions and symmetry properties of $F$-s come from
the generalized $\theta$-pe\-riodicity (\ref{pe}). We shall
examine the case of integer $n$, half-integer case is similar.

Applying (\ref{pe}) twice we obtain restrictions dependent on $n$:
\bea
\mbox{even}~~ n
,~~~~~~~~ n=2k
&&\Rightarrow F^0 =0 ,\nonumber\\
\mbox{odd}~~ n
&& \Rightarrow F_0 =F^+=F^- =0 .\nonumber
\eea

Applying (\ref{pe}) once we get that only the following functions $F$ are
consistence with the general periodicity:
\bea
n=4k:~~~~&& F^0 =0\nonumber\\
         && F_0(f,g) =F_0(g,f)  \nonumber\\
         && F^+(f,g) =F^+(g,f)  \nonumber\\
         && F^-(f,g) =-F^-(g,f) ,
\label{4k}
\eea
\bea
n=4k+2:&& F^0 =0\nonumber\\
         && F_0(f,g) =-F_0(g,f)  \nonumber\\
         && F^+(f,g) =-F^+(g,f)  \nonumber\\
         && F^-(f,g) =F^-(g,f)   ,
\label{4k+2}
\eea
\bea
n=4k+1:&& F_0 =F^+=F^-=0\nonumber\\
             && F^0 (f,g)=F^0 (g,f) ,
\label{4k+1}
\eea
\bea
n=4k+3:&& F_0 =F^+=F^-=0\nonumber\\
             && F^0 (f,g)=-F^0 (g,f) .
\label{4k+3}
\eea

\subsection{Hamiltonian}

We deduce the explicit expression for $H_B$ in new variables as
follows. At first, we calculate the metric tensor
$g^{\alpha\beta}$: $\delta^{ab}\delta_{ij} d\phi^a_i d\phi^b_j
=g^{\alpha\beta}dx^\alpha dx^\beta $ with $\vec x
=(f,g,\theta,\gamma_1 ,\gamma_2 ,\alpha)$. At second, we obtain
the Hamiltonian as the Laplace-Beltrami operator for the metric
$g^{\alpha\beta}$ plus the potential term
 $$\frac{1}{8g_s} f^2
g^2.
$$
 This straightforward but rather tedious calculation was
performed with partial use of $~~~$ "Maple". The result is the
following: {\arraycolsep=0mm
  \bea
 H_B&=&-g_s
 \left\{
  \Delta + D -
\frac{A}{a^2}(\vec{l})^2+ \frac1{(b)^2}
\left(
c
\frac{\partial^2}{\partial\theta^2} + 4a
\frac{\partial}{\partial\gamma_1} \frac{\partial}{\partial\theta}
\right)
 -\frac{b\sin 2\gamma_1}{a^2 \sin\alpha}
 \left( \cos\alpha
\frac{\partial}{\partial\gamma_1} -
\frac{\partial}{\partial\gamma_2}
 \right)
\frac{\partial}{\partial\alpha}
  \right.
   \nonumber\\
  &
 +& \frac{b \cos2\gamma_1}{a^2}
\frac{\partial^2}{\partial\alpha^2} - \frac{B}{a^2 b^2}
\frac{\partial^2}{\partial\gamma_1^2}
  \left.
+  \frac{b\sin 2\gamma_1}{2a^2 \sin^2\alpha}
\left(
(1+\cos^2\alpha)
\frac{\partial}{\partial\gamma_1} -
2\cos\alpha\frac{\partial}{\partial\gamma_2}
\right)
\right\}
+\frac1{8g_s}a^2 ,
 \label{hnew}
 \eea
}
  here $\Delta$ is Laplacian in
$f$, $g$ and:
$$
 D= \frac{1 }{J} \left( \frac{\partial J}{\partial
f} \frac{\partial }{\partial f}+ \frac{\partial J}{\partial g}
\frac{\partial }{\partial g} \right),
 $$
$$
a=f g,~~b=f^2 - g^2
,~~ c=f^2 +g^2 , ~~ J=ab ,
$$
$$
A=f^2\sin^2\gamma_1
+g^2\cos^2\gamma_1,
$$
$$
 B=f^6\sin^2\gamma_1-3f^4
g^2\sin^2\gamma_1 -3f^2g^4\cos^2\gamma_1 +g^6\cos^2\gamma_1.
$$

Note that there are terms with the first order derivatives
$\frac{\partial}{\partial f}$ and $\frac{\partial}{\partial g}$
collected in $D$. Although the coefficients for these terms are
functions of $f$, $g$ only, these terms are absent in the usual
quantization of the two-dimensional toy model  \cite{BMS}, as in
the pure bosonic case as in the fermionic one\footnote{The
appearance of the first order derivatives upon the correct
quantization was at first observed in the pioneering paper
\cite{Sav}}. These first order derivatives in couple with the zoo
of angular dependent terms in (\ref{hnew}) radically modify the
potential.

\subsection{Shr\"odinger Equation for Harmonics}

Now we substitute the harmonic expansion for the wave function
$\Psi_n$ into the system (\ref{schrod}) with the explicit
Hamiltonian $H_B$ given in (\ref{hnew}). The result is:

for {\bf $n$ - odd} (only $F^0\ne 0$) we get one PDE in $f$ and
$g$
\be
-g_s
\left\{
\Delta +D
-(n-1)^2\frac{c}{b^2}-\frac{c}{a^2}
\right\}
F^0+\frac{a^2}{8g_s}F^0
 = EF^0,
\label{eq4s0}
\ee

for {\bf $n$ - even} we get a system of three PDE in $f$ and
$g$
\bea
&&
-g_s
\left\{
\Delta +D
-n^2 \frac{c}{b^2}
\right\}
F_0
+\frac1{8g_s}a^2F_0
-\frac{f+g}{\sqrt{2}}F^++\frac{f-g}{\sqrt{2}}F^-
=EF_0
\label{eq1s}\\
&&
-g_s
\left\{
\left(
\Delta +D
-(n-1)^2\frac{c}{b^2}-(n-1)\frac{4a}{b^2}
-\frac{(f^4+g^4)c}{2a^2 b^2}
\right)
\right.
F^+
\left.
-\frac{b}{2a^2}
F^-
\right\}
\nonumber \\
 &&
\hspace{7cm}
 +\frac{a^2}{8g_s}F^+
 -F_0\frac{f+g}{2\sqrt{2}}
=EF^+
\label{eq4sp}\\
&&-g_s
\left\{
\left(
\Delta +D
-(n-1)^2\frac{c}{b^2}+(n-1)\frac{4a}{b^2}
-\frac{(f^4+g^4)c}{2a^2 b^2}
\right)
F^-
-\frac{b}{2a^2}F^+
\right\}\nonumber \\
&&\hspace{7cm}
+
\frac{a^2}{8g_s}F^-+F_0\frac{f-g}{2\sqrt{2}}
=EF^- .
\label{eq4sm}
\eea

\section{The Spectrum}

Remind, that for any $E\ne 0$ eigenfunctions are combined into the
quartets $V_{E,n_q}$ with quantum numbers $n_q, n_q-1/2, -n_q+2,
-n_q+3/2$. To examine the spectrum it is sufficient to search for
any one of these four functions constrained by the additional
constraints coming from supercharges $Q$ and $\bar{Q}$. Next we
search for differential equations for these constraints.

\subsection{Supercharges }

In the basis (\ref{vector}) the supercharges $Q$ and $\bar Q$ take
the form of anti-block-diagonal matrices:
\be
\nonumber
Q=\left(
\begin{array}{ll}
0&Q_+^*\\
Q_-&0
\end{array}
\right)
~~~~~~~~
\bar{Q}=\left(
\begin{array}{ll}
0&Q_-^*\\
Q_+&0
\end{array}
\right),
\ee
where asterisk stands for Hermitian conjugation.

The $4\times 4$ blocks are calculated to give:
$$
Q_-=\frac{-1}{\sqrt{2}}
\left(
\begin{array}{cccc}
0&\sqrt{ g_s}\pi^1&\sqrt{ g_s}\pi^2&\sqrt{ g_s}\pi^3\\
\sqrt{ g_s}\pi^1&0&\frac{-i}{\sqrt{ g_s}} {\omega}^3&\frac{i}{\sqrt{
g_s}}{ \omega}^2\\
\sqrt{ g_s}\pi^2& \frac{i}{\sqrt{ g_s}}{ \omega}^3&0&\frac{-i}{\sqrt{
g_s}} {\omega}^1\\
\sqrt{ g_s}\pi^3& \frac{-i}{\sqrt{ g_s}} {\omega}^2&\frac{i}{\sqrt{
g_s}} {\omega}^1&0
\end{array}
\right),
$$
$$
Q_+=\frac{-1}{\sqrt{2}}
\left(
\begin{array}{cccc}
0&
\frac{i}{\sqrt{ g_s}} \omega^1&  \frac{i}{\sqrt{ g_s}} \omega^2&
\frac{i}{\sqrt{ g_s}} \omega^3\\
\frac{i}{\sqrt{ g_s}} \omega^1&0& \sqrt{ g_s}\bar\pi^3&-\sqrt{
g_s}\bar\pi^2\\
\frac{i}{\sqrt{ g_s}} \omega^2& - \sqrt{ g_s}\bar\pi^3&0&  \sqrt{
g_s}\bar\pi^1\\
\frac{i}{\sqrt{ g_s}} \omega^3&  \sqrt{ g_s}\bar\pi^2&-  \sqrt{
g_s}\bar\pi^1&0\\
\end{array}
\right);
$$
here we used the complex notation (\ref{com}),
$\vec{\pi}=\vec{\pi}_1-i\vec{\pi}_2$
and
$\vec{\omega}=\vec{\varphi_1}\times\vec{\varphi_2}$.

For integer $n$ (see eq.(\ref{nn})):
$$
\Psi\equiv
\left(
\begin{array}{rcc}
&\psi_0&\nonumber\\
&\vec{\psi}&\nonumber\\
&0&\nonumber\\
&\vec 0& \nonumber
\end{array}
\right) ~~~
$$
and equations $Q\Psi =0 $, $\bar{Q}\Psi =0$ are reduced to
{\arraycolsep=0mm
\begin{eqnarray}
 Q_-
\left(
\begin{array}{c}
\psi_0\\
\vec{\psi}
\end{array}
\right)
&
{\displaystyle
=  \frac{-1}{\sqrt{2}}
}
\left(
\begin{array}{l}
\sqrt{ g_s}(\vec{\pi}\vec{\psi})\\
\sqrt{ g_s}\vec{\pi}\psi_0
+\frac{i}{\sqrt{ g_s}}\vec{\omega}\times\vec{\psi}
\end{array}
\right)
&
= \frac{-1}{\sqrt{2}}
\left(
\begin{array}{l}
\sqrt{ g_s}\mbox{div}\vec{\psi}\\
\sqrt{ g_s}\mbox{grad}\psi_0
+\frac{i}{\sqrt{ g_s}}\vec{\omega}\times\vec{\psi}
\end{array}
\right) =0 .
\label{q-}
\\
&&\nonumber\\
Q_+
\left(
\begin{array}{c}
\psi_0\\
\vec{\psi}
\end{array}
\right)
&
{\displaystyle
 = \frac{-1}{\sqrt{2}}
}
\left(
\begin{array}{l}
\frac{i}{\sqrt{ g_s}}(\vec{\omega}\vec{\psi})\\
\frac{i}{\sqrt{ g_s}}\vec{\omega}\psi_0
-\sqrt{ g_s}\vec{\bar{\pi}}\times\vec{\psi}
\end{array}
\right)
&
= \frac{-1}{\sqrt{2}}
\left(
\begin{array}{l}
\frac{i}{\sqrt{ g_s}}(\vec{\omega}\vec{\psi})\\
\frac{i}{\sqrt{ g_s}}\vec{\omega}\psi_0 -
\sqrt{ g_s}\bar{\mbox{rot}}\vec{\psi}
\end{array}
\right) =0 .
\label{q+}
\end{eqnarray}
}

\subsection{Supercharges in $SU(2)\times SO(2)$ Parametrisation}

At first we do not specify the parity of $n$ and
 employ the harmonics expansion (eqs. (\ref{sol3}), (\ref{sol+}), (\ref{sol-}))
for $\Psi_n$. The nondifferential terms in
 (\ref{q-}), (\ref{q+}) can be easily calculated to give:
\be
\vec{\omega}=\frac{fg}2\vec{v},
\ee
\be
(\vec\omega \vec\psi)=e^{i(n-1)\theta}\frac{fg}2F^0,
\label{omps}
\ee
\be
\vec\omega \times \vec\psi = -e^{i(n-1)\theta}\frac{fg}2
\left(e^{i\gamma_1}F^+ \vec{u}
+e^{-i\gamma_1}F^- \vec{w}\right)
 \label{om}
\ee
where we have introduced three vectors:
\bea
\vec{u}&=&\{\cos\gamma_2+i\cos\alpha\sin\gamma_2,~~
\sin\gamma_2-i\cos\alpha\cos\gamma_2,~~ -i\sin\alpha\},\nonumber\\
\vec{v}&=&\{\sin\alpha\sin\gamma_2,~~ -\sin\alpha\cos\gamma_2,~~
\cos\alpha\},\nonumber\\
\vec{w}&=&\{\cos\gamma_2-i\cos\alpha\sin\gamma_2,~~
\sin\gamma_2+i\cos\alpha\cos\gamma_2,~~ i\sin\alpha\}.
\eea

 The explicit expressions for $\vec \pi$ $\vec{\bar{ \pi}}$,
are too bulky and are listed in the Appendix. Here we quote the
results deduced by means of the computer calculations:
\be
\vec{\pi}\psi_0 = \frac1{i\sqrt{2}}e^{i(n-1)\theta}
\left(
 e^{i\gamma_1}
R^+_n
 F_0 \vec{u} +
 e^{-i\gamma_1}
R^-_n
 F_0 \vec{w}
 \right) ,
 \label{grad}
\ee
\be
\vec{\pi}\vec\psi
 =\sqrt{2}e^{i(n-2)\theta}
\left[
\left(
R^-_n -
\frac{(f-g)}{fg}
\right)
F^+
\right .
\left. -
\left(
R^+_n +
\frac{(f+g)}{fg}
\right)
F^-
\right],
\label{pi}
\ee
\be
\vec{\bar{\pi}}\times\vec{\psi}= -i2\sqrt{2}e^{in\theta}\frac1{fg}
\left(
 R^+_{2-n}
 F^+ +
 R^-_{2-n}
F^-
\right)
\vec{\omega}
+LF^0,
\ee
where
\be
R^+_n=
\left(
\frac{\partial}{\partial_f} +\frac{\partial}{\partial_g} +\frac n {f+g}
\right),~~~
R^-_n=
\left(
\frac{\partial}{\partial_f} -\frac{\partial}{\partial_g} +\frac n {f-g}
\right)
\ee
and$L$ is an irrelevant differential operator

Remind, that the harmonics content of the wave function $\Psi_n$ differs
with respect to the parity of $n$:
\bea
\mbox{for odd}~~ n: && F_0 =F^+ =F^- =0 \nonumber \\
\mbox{for even}~~ n: && F^0 =0 \nonumber
\eea

At the same time it is seen from eqs. (\ref{om}), (\ref{pi}) that
all the terms in eq.(\ref{q-})
for $Q_-$ are expressed solely in terms of $F^\pm$ and $F_0$.
Hence, for
\be
\mbox{\bf odd}~~n:~~~~~~ Q\Psi_{En} \equiv 0.
\label{A}
\ee
On the other hand, it is known that  $H\Psi_n =E\Psi_n$ and $Q\Psi_n =0$
implies
$\bar{Q}\Psi_n \ne 0$, i.e. for
\be
\mbox{\bf odd}~~n:~~~~~~ \bar{Q}\Psi_{En} \ne 0.
\label{B}
\ee

The case of even $n$ is more involved. By using the above formulae we get
\be
\begin{array}{rcl}
\mbox{\bf even }~n:~
\bar{Q}\Psi =0
&\rightarrow&
\sqrt{2g_s}
\left(
\displaystyle
R^+_{2-n} F^+ + R^-_{2-n}F^-
\right)
\displaystyle
+ \frac {fg}{2\sqrt{g_s}}F_0 =0
\end{array}
\label{qbar}
\ee
and
\be
\begin{array}{rcl}
\mbox{\bf even}~n:~
Q\Psi =0
&\rightarrow&
\left\{
\begin{array}{l}
\sqrt{2g_s}
R^+_n F_0
\displaystyle
+\frac{fg}{\sqrt{g_s}}F^+ =0
\\
\\
\sqrt{2g_s}
R^-_n F_0
\displaystyle
+\frac{fg}{\sqrt{g_s}}F^- =0
\\
\\
\left(
R^-_n -
\displaystyle
\frac{(f-g)}{fg}
\right)
F^+ -
\left(
\displaystyle
R^+_n -
\frac{(f+g)}{fg}
\right)
F^- =0 . \\
\end{array}
\right.
\end{array}
\ee
 One can express $F^+$ and $F^-$ from the first and the second
equations respectively. The third equation of the system above is
nothing but the consistency condition for the first and the second
ones.

Now we are in position to examine the restrictions on $V_{E,n_q}$
coming from $Q\Psi_n =0$ and $\bar{Q}\Psi_n =0$. As it was shown
above $n$ can be integer or half-integer. We start with the case
of integer $n_q$. (We omit the subindex $q$ below). The content of
$V_{En}$ is the following: $\Psi_{n}$, $\Psi_{n-\frac12}$,
$\Psi_{-n+2}$ and $\Psi_{-n+\frac32}$ with $\bar{Q}\Psi_{n}=0$ and
$\bar{Q}\Psi_{-n+2}=0$. Both $n$ and $-n+2$ have the same parity
so according to (\ref{B}) odd $n$ are forbidden. For $n$ even the
constraint $\bar{Q}\Psi_n =0$ results in PDE (\ref{qbar}). We
shall examine this case in Sec. 5.4.

Next we turn to a half-integer $n=n' +\frac12$. $V_{En'+\frac12}$ is spanned
by $\Psi_{n'+\frac12}$, $\Psi_{n'}$, $\Psi_{-n'+\frac32}$ and $\Psi_{-n'+1}$
with $Q\Psi_{n'} =0$, $Q\Psi_{-n'+1}=0$. This time one of two numbers either $n'$
or $-n'+1$ will be odd. Let, for instance, $n'$ to be odd. As follows from
(\ref{A}) $Q\Psi_{n'} =0$ and the only equation to be solved is the
Shr\"odinger
equation (\ref{eq4s0}). It will be examined in the next section.

\subsection{Half-integer $n_q$}

In this section we examine the solution of eq. (\ref{eq4s0})
$$
-g_s \left( \Delta +D
-\frac{g^2+f^2}{g^2f^2}-(n-1)^2\frac{g^2+f^2}{(f^2-g^2)^2} \right)
F^0+\frac{f^2g^2}{8g_s}F^0=EF^0
$$
by using the Born-Oppenheimer
method in the region $f\gg g$, as it is adopted in current
literature. This procedure yields an approximate formulae for the
spectrum and wave function. Note, that in the current variables
$\|\Psi\|$ is proportional to $\int \bar{F^0} F^0 J dfdg$.

To eliminate the first derivatives term $D$ we search for the
wave function in the form
$$
F^0=\frac 1{\sqrt{J}}\varphi^0
$$
that yields the equation

\be
-g_s
\left(
\Delta+\frac 14\frac{(f^2+g^2)^3}{f^2g^2(f^2-g^2)^2}
-\frac{g^2+f^2}{g^2f^2}-(n-1)^2\frac{g^2+f^2}{(f^2-g^2)^2}
\right)
\varphi^0+\frac{f^2g^2}{8g_s}\varphi^0=E\varphi^0
\label{oddr}
\ee
and
$$
\|\Psi\|=C\int\bar{\varphi}^0\varphi^0 dfdg .
$$

We start with introducing an extra parameter $a$ as follows: $f\to
f/a^2$ and $g\to ag$ and expanding over the powers of $a$.
Separating the variables as $\varphi^0 = \psi_{m}(g|f)\chi_k(f)$
we get the leading term as
$$
\left
(-\frac{\partial^2}{\partial g^2}+\frac34\frac{1}{g^2}
+b^2g^2
\right)
\psi_m=\varepsilon_m\psi_m
$$
where $b=\frac f{2\sqrt{2}g_s}$.

The discrete spectrum is given by
$\varepsilon_m=4b(m+1)$ where $m=0,1,2,\dots$ and
$\psi_m$ is expressed in terms of Laguerre polynomials
$$
\psi_m=(-1)^m\frac 1{\sqrt{g}}bg^2e^{-\frac 12bg^2}L_m^1(bg^2) .
$$

The next to leading term in the $a$-expansion
yields the equation for $\chi$ (we omit
the index $k$)
$$
(-\frac{\partial^2}{\partial f^2}+4\frac
f{2\sqrt{2}g_s}(m+1))\chi=E\chi
 $$
 where we temporary suppressed
the term $1/f^2$. The change of variables
$f=x+E\frac{\sqrt{2}g_s}{2(m+1)}$ yields an equation for Airy
function
$$
\left(\frac{\partial^2}{\partial x^2}-\frac
{\sqrt{2}(m+1)}{g_s}x\right)\chi=0.
$$
The solution in terms of $y=(\frac {\sqrt{2}(m+1)}{g_s})^{1/3}x$ is
$$
\chi=\left\{
\begin{array}{lll}
\sqrt{\frac y{3\pi}}K_{1/3}\left(\frac 23y^{3/2}\right)&~\mbox{for}~&y>0\\
\frac {\sqrt{\pi|y|}}3
\left(J_{-1/3}\left(\frac 23|y|^{3/2}\right)+
J_{1/3}\left(\frac 23|y|^{3/2}\right)\right)
&~\mbox{for}~&y<0 .\\
\end{array}
\right.
$$
Imposing a boundary condition $\chi(0)=0$ we get
$$
J_{-1/3}(\rho_k)+J_{1/3}(\rho_k)=0
$$
where
$$
\rho_k=\frac{\sqrt{2}}3\frac{g_s}{m+1}E^{3/2} .
$$
By using
$$
J_{\nu}(\rho)+J_{-\nu}(\rho)=2\cos\frac {\pi\nu}2
\left(J_{\nu}(\rho)\cos\frac{\pi\nu}2-N_{\nu}(\rho)\sin\frac{\pi\nu}2\right)
$$
and asymptotic of $J(\rho)$ and $N(\rho)$ for large $\rho$
\bea
J_{\nu}(\rho)\approx \sqrt{\frac 2{\pi \rho}}\cos\left(\rho-\frac
{\pi}2\nu-\frac{\pi}4\right)\nonumber\\
 N_{\nu}(\rho)\approx
\sqrt{\frac 2{\pi \rho}}\sin\left(\rho-\frac
{\pi}2\nu-\frac{\pi}4\right)\nonumber
\eea
we get
$$
\rho_k=\pi k+\frac34\pi,~~~k=0,1,2,\dots
$$

Thus, the result is
$$
E_{m,k}=\left(\pi\frac 3{\sqrt{2}}\frac{(m+1)(k+\frac34)}{g_s}\right)^{2/3}
$$
and
\be
\varphi_{m,k}\!\sim\! c_{m,k}fg^{3/2}\xi
e^{-\frac{\alpha}2 fg^2}L_m^1(\alpha fg^2)
\times
\left\{
\begin{array}{ll}
\frac {\sqrt{3}}{\pi}
K_{1/3}\!\left(\frac 23a^{1/2}\xi^3\right),
&f>\frac 1a E_{m,k}\\
J_{1/3}\!\left(\frac 23a^{1/2}\xi^3\right)+
J_{-1/3}\!\left(\frac 23a^{1/2}\xi^3\right),
&f<\frac 1a E_{m,k}
\end{array}
\right.
\label{psi}
\ee
where
$$
\alpha=\frac 1{2\sqrt{2}g_s},~~~
a=\frac{\sqrt{2}(m+1)}{g_s},~~~
\xi=\left|f-\frac 1a E_{m,k}\right|^{1/2}
$$

The first correction to the energy is given by
$$
\Delta E=\frac{\int\bar\chi V\chi df}{\int\bar\chi\chi df} .
$$
In our case $V=\frac{(n-1)^2-1/4}{f^2}$ and the direct calculation gives
$$
E=\left(\frac{3\pi}{\sqrt{2}}\right)^{2/3}
\left(\frac{(m+1)(k+\frac 34)}{g_s}\right)^{2/3}
\left(1+\frac43\frac{(n-1)^2-\frac14}{(k+\frac34)^2}\right) .
$$

\subsection{Even $n_q$ and continuous spectrum}

Here we apply the Born-Oppenheimer method in the region $f\gg g $
to the system (\ref{eq1s}) - (\ref{eq4sm}). Our aim is to employ
the toy-model-like ansatz  for the wave function and justify the
appearance of the continuous spectrum \cite{WLN}.

For $f\gg g$ the leading part in the system (\ref{eq1s}) -
(\ref{eq4sm}) reads
\bea
&&
-(
\Delta +D)\tilde F_0
+\frac1{8g^2_s}a^2\tilde F_0
-\frac{f}{2g_s}F^++\frac{f}{2g_s}F^-
=E\tilde F_0
\label{1b}\\
&&
-(\Delta +D
-\frac{1}{2g^2})F^+ +\frac{a^2}{8g^2_s}F^+
+\frac{1}{2g^2}F^-
 -\frac{f}{2g_s}\tilde F_0
=EF^+
\label{2b}\\
&&-(\Delta +D
-\frac{1}{2g^2})F^- +
\frac{a^2}{8g^2_s}F^-
+\frac{1}{2g^2}F^+
+\frac{f}{2g_s}\tilde F_0
=EF^-
\label{3b}
\eea
where we rescaled $F_0 =\sqrt{2}\tilde F_0$.

Taking $F^- =- F^+$ we find that (\ref{2b}) and (\ref{3b})
coincide. The reduced system reads:
 \bea
&& -( \Delta +D)\tilde F_0 +\frac1{8g^2_s}a^2\tilde F_0
-\frac{f}{g_s}\tilde F_0 =E\tilde F_0 \label{1c}\\
&&
-(\Delta +D
) F^+ +\frac{a^2}{8g^2_s} F^
 -\frac{f}{2g_sA}\tilde F_0
=E F^+ .
\label{2c}
\eea
It is obvious that (\ref{1c}) and
(\ref{2c}) will coincide as well if one put $F^+
=\frac1{\sqrt{2}}\tilde F_0$.

In the asymptotic region the rescaling yields $D\to 1/(4g^2)$ and
we are left with the equation
$$
-( \Delta +\frac 1{4g^2})\tilde F_0 +\frac1{8g^2_s}f^2g^2\tilde
F_0 -\frac{f}{\sqrt{2}g_s}\tilde F_0 =E\tilde F_0
$$
Recall, that equation
$$
-( \frac{\partial}{\partial g^2} +\frac 1{4g^2})\psi_m
+\frac1{8g^2_s}f^2g^2\psi_m =\varepsilon_m\psi_m
 $$
 has
normalisable solutions for
$\varepsilon_m=\frac{\sqrt{2}f}{g_s}(m+1/2)$, where $1/2$ is very
important (not $1$ !).

The next step is to solve the equation
$$
-( \frac{\partial}{\partial f^2}+O(\sim1/f^2))\chi=E\chi .
$$

We confirm that for $m=0$ the linear term from the "energy"
$\varepsilon_m$ is fully cancelled by the linear potential
$\frac{f}{\sqrt{2}g_s}$ thus producing the continuous spectrum.

\section{Discussion and Conclusion}

 The results obtained in Section 5 are collected in
the table presented in Fig.3.  States from the quartets are
situated along the horizontal lines. Eigenvalues of $N$ for the
states in a quartet $n_q$ are $n_q, n_q-1/2$ and $-n_q+2,
-n_g+3/2$. Each item of the quartet can be obtained from any other
one by the action of supercharges or the permutation operator
$P$. Our results show that odd $n_q$ are forbidden, for even $n_q$
the spectrum has a continuous part as well as a discrete one,
meanwhile for half-integer $n_q$ the spectrum is purely discrete.
\begin{figure}[h]
\begin{center}
\epsfig{file=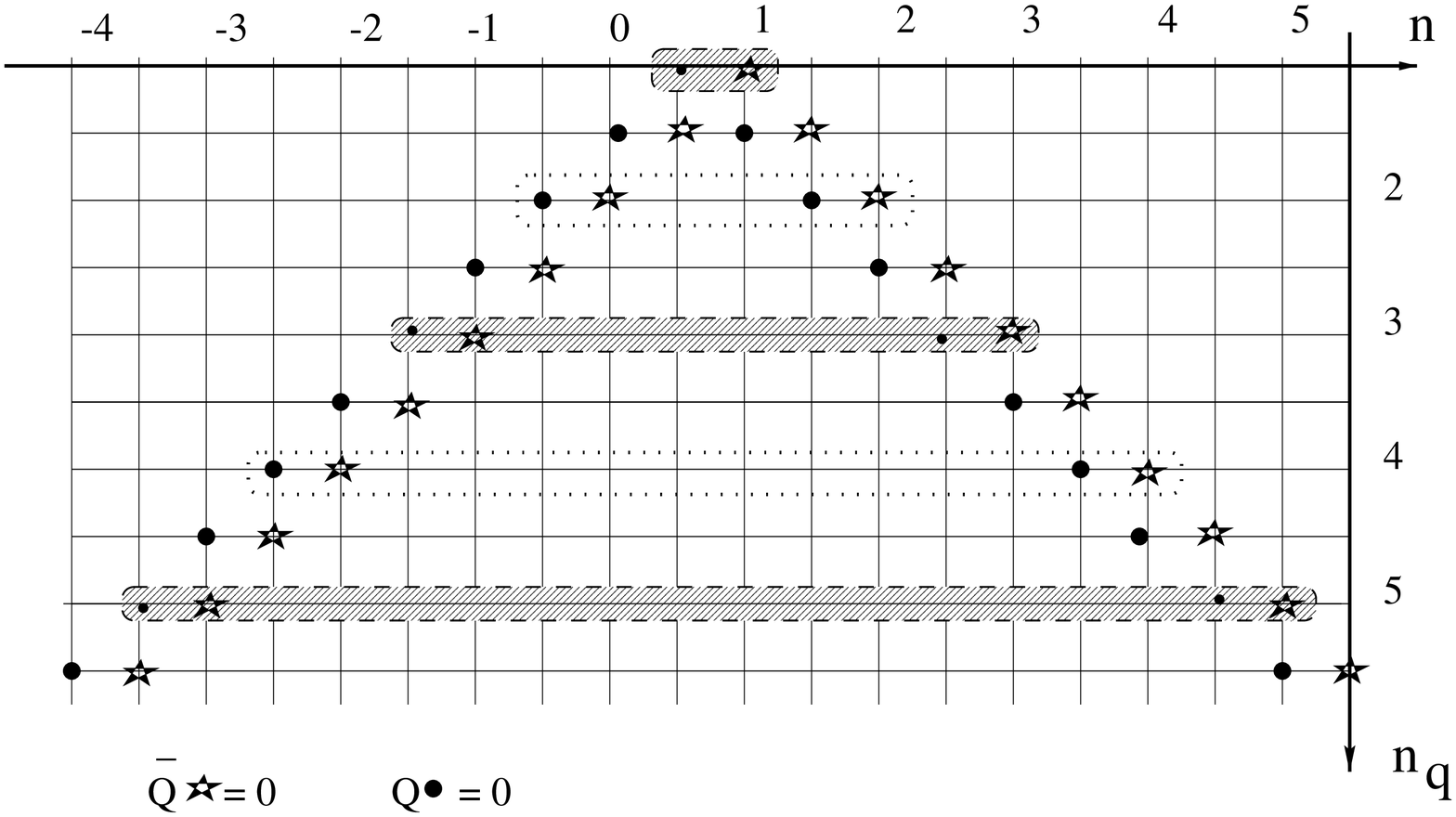,
   width=450pt,
  angle=0
 }
\end{center}
\label{Fig1} \caption{Structure of the spectrum:  stars correspond
to $\bar {Q}\Psi=0, ~~$filled circles correspond to $Q\Psi =0$;
domains bounded by  dots  correspond to the presence of the
continuous spectrum; shared domains correspond to excluding values
of $n$ and $n_q$. Quartets are located along the horizontal lines.
}
\end{figure}

Table 3 demonstrates that there is a possibility to put a
superselection rule to exclude the presence of the continuous
spectrum. Namely, taking any state from quartet with half-integer
$n_q$ one gets with guarantee the state of discrete spectrum. One
can also specify the purely discrete sector by using the
supercharges, i.e. the states with integer $n$ satisfy the
constraint $Q\Psi_n=0$, meanwhile states with half-integer $n$
satisfy $\bar{Q}\Psi_n=0$.

 Besides the continuous spectrum $SU(2)\times
SO(2)$ matrix supersymmetric quantum mechanics possesses one more
distinguish feature as compared with the bosonic counterpart. This
concerns the behaviour of the wave functions in half-integer
$n_q$ sector, where spectrum is purely discrete. The expression
(\ref{psi}) for the wave function illuminates this difference. It
is not a difficult task to realize that in the pure bosonic case
the maximum of the wave function appears at the minimum of the
classical potential $(fg)^2$, i.e. at the bottom of the valley
$f=0$ as it should be by the semiclassical reasons. For the
supersymmetric case the picture is quite different, the classical
equilibrium line have nothing to do with the maximum of the wave
function (\ref{psi}), this time the wave function tends to zero as
$f\to 0$. The origin of this transformation  is the reflecting
wall like $\frac1{f^2}$ created by the activation of the $SU(2)$
angular degrees of freedom for the triplet $(l=1)$ solution.

Let us note that we have obtained only the asymptotic  formula for
spectrum. Numeric investigations (see \cite{Sal} and references
therein) allow to trust the asymptotic formula. It would be
interesting to perform numerical calculations also in our case. It
would be also interesting to study $\alpha '$-corrections
\cite{AFK} to the spectrum.

It was  argued \cite{AMRV} that the holographic feature of the
matrix theory can be related with the repulsive feature of energy
eigenvalues in the quantum chaotic system. Relation between chaos
and holography has been discussed recently in \cite{thooft}.
Quantum chaos in supersymmetric matrix quantum mechanics will be a
subject of future investigationss.

\section*{Acknowledgments}

We would like to thank L.O. Chekhov, B.V. Medvedev, A.K.
Pogrebkov, O.A. Rytchkov, N.A. Slavnov and I.V. Volovich for
useful discussions. This work was supported in part by by RFFI
grant 99-01-00166  and by grant for the leading scientific schools
96-15-96208. I.A. is also supported by INTAS grant 96-0698.
\newpage
\section{Appendix}

Derivatives in new coordinates are
\bea
\frac{\partial}{\partial\varphi^+}&=&\frac{\sqrt{2}}4e^{-i(\theta+\gamma_2)}\left(\frac{\sin\alpha(f\cos\gamma_1+ig\sin\gamma_1)}{fg}\frac{\partial}{\partial\alpha}+
\frac{(f\sin\gamma_1-ig\cos\gamma_1)}{fg}\frac{\partial}{\partial\gamma_2}-\right.\nonumber\\
&&\frac{(f^2g\sin\gamma_1+f^3\cos\alpha\sin\gamma_1+
ig^3\cos\alpha\cos\gamma_1+ifg^2\cos\gamma_1)}{(fg(f^2-g^2))}\frac{\partial}{\partial\gamma_1}-\nonumber\\
&&\frac{(g\sin\gamma_1+f\cos\alpha\sin\gamma_1+if\cos\gamma_1+ig\cos\gamma_1\cos\alpha)}{f^2-g^2}\frac{\partial}{\partial\theta}-\nonumber\\
&&\left.(i\cos\alpha\sin\gamma_1-\cos\gamma_1)\frac{\partial}{\partial
f}- (\cos\alpha\cos\gamma_1-i\sin\gamma_1)\frac{\partial}{\partial
g}\right)\\
\frac{\partial}{\partial\varphi^-}&=&-\frac{\sqrt{2}}4e^{-i(\theta-\gamma_2)}\left(\frac{\sin\alpha(f\cos\gamma_1+ig\sin\gamma_1)}{fg}\frac{\partial}{\partial\alpha}+
\frac{(f\sin\gamma_1-ig\cos\gamma_1)}{fg}\frac{\partial}{\partial\gamma_2}-\right.\nonumber\\
&&\frac{(-f^2g\sin\gamma_1+f^3\cos\alpha\sin\gamma_1+
ig^3\cos\alpha\cos\gamma_1-ifg^2\cos\gamma_1)}{(fg(f^2-g^2))}\frac{\partial}{\partial\gamma_1}+\nonumber\\
&&\frac{(g\sin\gamma_1-f\cos\alpha\sin\gamma_1+if\cos\gamma_1-ig\cos\gamma_1\cos\alpha)}{f^2-g^2}\frac{\partial}{\partial\theta}-\nonumber\\
&&\left.(i\cos\alpha\sin\gamma_1+\cos\gamma_1)\frac{\partial}{\partial
f}- (\cos\alpha\cos\gamma_1+i\sin\gamma_1)\frac{\partial}{\partial
g}\right)\\
\frac{\partial}{\partial\bar{\varphi}^+}&=&-\frac{\sqrt{2}}4e^{i(\theta-\gamma_2)}\left(\frac{\sin\alpha(f\cos\gamma_1-ig\sin\gamma_1)}{fg}\frac{\partial}{\partial\alpha}+
\frac{(f\sin\gamma_1+ig\cos\gamma_1)}{fg}\frac{\partial}{\partial\gamma_2}-\right.\nonumber\\
&&\frac{(-f^2g\sin\gamma_1+f^3\cos\alpha\sin\gamma_1-
ig^3\cos\alpha\cos\gamma_1+ifg^2\cos\gamma_1)}{(fg(f^2-g^2))}\frac{\partial}{\partial\gamma_1}-\nonumber\\
&&\frac{(-g\sin\gamma_1+f\cos\alpha\sin\gamma_1+if\cos\gamma_1-ig\cos\gamma_1\cos\alpha)}{f^2-g^2}\frac{\partial}{\partial\theta}+\nonumber\\
&&\left.(i\cos\alpha\sin\gamma_1-\cos\gamma_1)\frac{\partial}{\partial
f}+
(-\cos\alpha\cos\gamma_1+i\sin\gamma_1)\frac{\partial}{\partial
g}\right)\\
\frac{\partial}{\partial\bar{\varphi}^-}&=&\frac{\sqrt{2}}4e^{i(\theta+\gamma_2)}\left(\frac{\sin\alpha(f\cos\gamma_1-ig\sin\gamma_1)}{fg}\frac{\partial}{\partial\alpha}+
\frac{(f\sin\gamma_1+ig\cos\gamma_1)}{fg}\frac{\partial}{\partial\gamma_2}+\right.\nonumber\\
&&\frac{(-f^2g\sin\gamma_1-f^3\cos\alpha\sin\gamma_1+
ig^3\cos\alpha\cos\gamma_1+ifg^2\cos\gamma_1)}{(fg(f^2-g^2))}\frac{\partial}{\partial\gamma_1}-\nonumber\\
&&\frac{(g\sin\gamma_1+f\cos\alpha\sin\gamma_1-if\cos\gamma_1-ig\cos\gamma_1\cos\alpha)}{f^2-g^2}\frac{\partial}{\partial\theta}+\nonumber\\
&&\left.(i\cos\alpha\sin\gamma_1+\cos\gamma_1)\frac{\partial}{\partial
f}- (\cos\alpha\cos\gamma_1+i\sin\gamma_1)\frac{\partial}{\partial
g}\right)\\
\frac{\partial}{\partial\varphi^3}&=&-\frac{\sqrt{2}}2e^{-i\theta}\left(\frac{\cos\alpha(if\cos\gamma_1-g\sin\gamma_1)}{fg}\frac{\partial}{\partial\alpha}+
\frac{\cos\alpha(if\sin\gamma_1+g\cos\gamma_1)}{fg\sin\alpha}\frac{\partial}{\partial\gamma_2}-\right.\nonumber\\
&&\frac{(f^2g\cos\gamma_1+if^3\cos^2\alpha\sin\gamma_1-
g^3\cos^2\alpha\cos\gamma_1-ifg^2\sin\gamma_1)}{(fg\sin\alpha(f^2-g^2))}\frac{\partial}{\partial\gamma_1}+\nonumber\\
&&\left.\frac{\sin\alpha(-g\cos\gamma_1+if\sin\gamma_1)}{f^2-g^2}\frac{\partial}{\partial\theta}-
\sin\alpha\sin\gamma_1\frac{\partial}{\partial
f}+\sin\alpha\cos\gamma_1\frac{\partial}{\partial g}\right)\\
\frac{\partial}{\partial\bar{\varphi}^3}&=&\frac{\sqrt{2}}2e^{i\theta}\left(\frac{\cos\alpha(if\cos\gamma_1+g\sin\gamma_1)}{fg}\frac{\partial}{\partial\alpha}+
\frac{\cos\alpha(if\sin\gamma_1-g\cos\gamma_1)}{fg\sin\alpha}\frac{\partial}{\partial\gamma_2}-\right.\nonumber\\
&&\frac{(-f^2g\cos\gamma_1+if^3\cos^2\alpha\sin\gamma_1+
g^3\cos^2\alpha\cos\gamma_1-ifg^2\sin\gamma_1)}{(fg\sin\alpha(f^2-g^2))}\frac{\partial}{\partial\gamma_1}+\nonumber\\
&&\left.\frac{\sin\alpha(g\cos\gamma_1+if\sin\gamma_1)}{f^2-g^2}\frac{\partial}{\partial\theta}+
\sin\alpha\sin\gamma_1\frac{\partial}{\partial
f}+\sin\alpha\cos\gamma_1\frac{\partial}{\partial g}\right) \eea


\end{document}